\documentclass{osa-article}

\journal{boe}

\usepackage{changes} 

\begin{document}

\title{Ultrasensitive THz Biosensor for PCR-free cDNA detection based on frequency selective surfaces}

\author{Christian Weisenstein,\authormark{1,*} Dominik Schaar,\authormark{2} Anna Katharina Wigger,\authormark{1} Heiko Sch\"afer-Eberwein,\authormark{1} Anja K. Bosserhoff,\authormark{2} and Peter Haring Bol\'{i}var\authormark{1}}

\address{\authormark{1}Institute of High Frequency and Quantum Electronics HQE, University of Siegen, Germany\\
\authormark{2}Institute of Biochemistry, Friedrich-Alexander-University Erlangen-N\"urnberg, Germany\\

\email{\authormark{*}christian.weisenstein@uni-siegen.de}} 


\begin{abstract}
THz technologies are a powerful tool for label-free detection of biomolecules. However, significant reduction of the lower detection limit is required to apply THz-sensors in biomedical diagnosis. This paper reports an ultrasensitive THz-biosensor based on asymmetric double split ring resonators (aDSRR) for the direct label- and PCR-free detection of DNA at physiologically relevant concentrations. We introduce selective functionalization and localized electric field concentration to enhance aDSRR sensitivity and specificity. The sensor characteristics are demonstrated using the human tumor marker MIA in cDNA samples produced from total RNA without PCR-amplification. Measurements of DNA samples with concentrations as low as $1.55\,\times\,10^{-12}\,$mol/l are presented.
\end{abstract}

\section{Introduction}
Recent substantial developments in genomic technologies have provided possibilities for widespread future health applications, e.g., predictive, preventive, and personalized medicine. These advances are closely coupled with the development of powerful, reliable, and efficient methods that can detect, analyze, and identify biomolecules, and assess their complex interactions in biological networks. Standard bioanalytical techniques for DNA analysis still rely on amplification techniques like polymerase chain reaction (PCR) in order to obtain higher detectable quantities and on fluorescently labeled DNA targets. Although sensitive and established, these techniques are time consuming and require extreme caution during preparation and analysis. There have been many attempts to develop biosensors using the so-called Lab-on-Chip technology as it promises to be a powerful, fast, and simple tool for DNA analysis \cite{Vo-Dinh2001,Liu2004,Wang2000}. However, the current methods still rely on fluorescence labeling with high system complexity. Fluorescence labeling, as well as PCR amplification, can modify the DNA strand configuration which can introduce diverse and unwanted interference with DNA samples, thereby corrupting the analysis \cite{Ozaki1992,Zhu1994,Zhu1997,Larramendy1998}.

Since resonances related to both macro- and bio-molecular interactions lie in the Terahertz (THz) frequency range, THz sensing and analysis of biomolecules has become an attractive alternative detection method. Vibration, torsion, and libration modes, as well as binding states cause resonant absorptions in the THz frequency range that result in characteristic material-specific spectral fingerprints, allowing for label-free THz analysis of biomolecules \cite{Froehlich1968,Zhuang1990}. The label-free sensing and analysis of biomolecules using THz radiation has already been demonstrated in the early 2000s \cite{BrucherseiferNagelBolivarEtAl2000,Markelz2000,Mickan2002,Fischer2002}, proving THz sensing to be a promising technique.

The relatively large wavelength of THz waves ($300\,\mu$m at $1\,$THz) compared to the size of biomolecules (typically $< 100\,$nm) strongly limits its sensitivity in terms of the minimum number of biomolecules that can be detected by a spectroscopic THz sensing approach. Frequency-selective surfaces (FSS) with a resonance frequency in the THz range may significantly enhance THz sensitivity \cite{DebusBolivar2007}. FSS are based on a periodic array of resonant metallic structures, designed to create a sharp resonance in the frequency response \cite{Fedotov2007a}. By applying a dielectric load to the FSS, the resonance frequency is detuned, resulting in a shift in resonance frequency compared to the unloaded FSS. In THz biosensing, this sensitive mechanism is used to detect biomolecules \cite{DebusBolivar2007,DebusAwadNagelEtAl2009}.

However, despite of such advances, the sensitivity of THz sensing techniques is still several orders of magnitude lower than state-of-the-art bioanalytical sensing techniques. THz techniques still require significant improvements in order to reach relevant detection sensitivities needed for real-world applications. Numerous publications on THz analyses have measured biomolecules with comparable molecular weights (MW) in aqueous solution and found that the sensitivity is strongly dependent on MW: as MW increases, the minimum detectable concentration decreases because heavier and larger biomolecules are detected more easily.

DNA composed of four 16-mer sequences with an MW of approx. $19527\,$g/mol, and a concentration of $4 \times 10^{-6}\,$mol/l can be measured in aqueous solution using THz spectroscopy \cite{Lvovska2010}, and for larger DNA strands with 133 base pairs (bp) with an MW of approx. $80942.1\,$g/mol a minimum concentration of $1.23 \times 10^{-9}\,$mol/l has been reported \cite{Arora2012}. At the protein level, bovine serum albumin (BSA) (MW of $66400\,$g/mol and concentration of $7.5 \times 10^{-5}\,$mol/l) has been detected with THz spectroscopy in aqueous solution \cite{Laurette2012}. In contrast, established bioanalytical tools like Enzyme-linked Immunosorbent Assays (ELISA) determine human serum albumin (HSA) (MW of $66437\,$g/mol) at concentrations as low as $3.16 \times 10^{-12}\,$mol/l, however by using signal amplification \cite{ELISA179887}. A recent publication applied THz measurements to a liver tumor marker using an FSS-based chip with a microfluidic structure for sample loading \cite{Geng2017}. Prior to the measurement, the authors removed the microfluidic channel from the FSS structure and dried the samples using N$_{2}$. Anti-alpha 1 fetoprotein antibody and antigen were measured with an MW of approx. $150000\,$g/mol and a concentration of $14.6 \times 10^{-9}\,$mol/l. This result demonstrates the sensitivity enhancement enabled by FSS-based chips in comparison to spectroscopic THz techniques. Despite of such advances, THz detection limits are still about a factor of 4000 less sensitive than other established bioanalytical techniques \cite{ELISA179887}.

In this paper, we present a newly developed highly sensitive THz biosensor based on selectively functionalized FSS for the detection of DNA and an experimental evaluation of its detection capabilities. The FSS sensing structure has been carefully designed to maximize sensitivity to enable measurements at physiologically relevant scales. Specificity to the target biomolecule is achieved by selective hybridization of complementary DNA (cDNA) strands. Selective functionalization of the FSS structure is used to chemically bind single-stranded (ss) oligo- or polynucleotide probe molecules to the biosensor surface.

High frequency excitation of the biosensor exhibits a strong concentration of the electric field at selectively functionalized resonant structures of the FSS. We measure synthetic and human DNA samples on our THz biosensor in ambient air using an all-electronic room temperature THz system. We provide experimental evidence of the high sensitivity of the THz biosensor using tumor markers produced from human RNA samples without requiring labeling nor a PCR amplification procedure.

\section{Biosensor and reader setup}
\subsection{Biosensor concept and design}
\begin{figure}[ht]
\centering
\includegraphics[width=\textwidth]{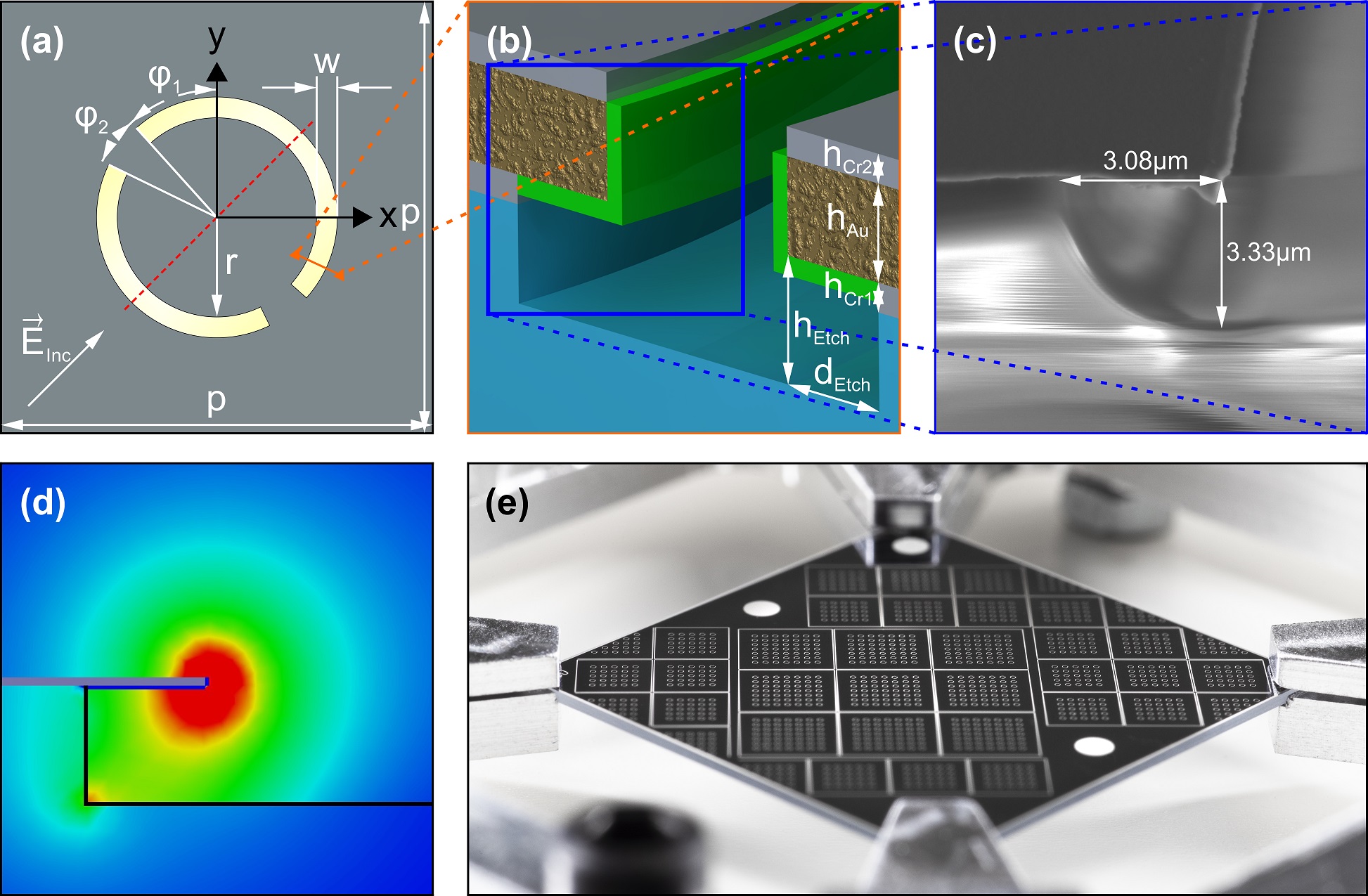}
\caption{(a) Schematic layout of the aDSRR structure and (b) cross section of one arc of the aDSRR (not to scale), showing the quartz substrate with the etched profile (blue) and the lithographic chromium layers (grey) enclosing the gold layer. The biofilm (green) is selectively functionalized on the open gold surfaces. (c) Cross-sectional SEM image of a fabricated biosensor with the undercut etched profile. (d) Simulated distribution of the electric field in the cross section of the aDSRR long arc. The maximum of the asymmetric E-field is concentrated at the edge of the free-standing metallic structure. (e) Complete biosensor with query fields consisting of aDSRR arrays of 5x5 and 7x7 elements.}
\label{fig:adsr_complete}
\end{figure}

The design of the biosensor is based on asymmetric double split ring resonator (aDSRR) arrays (Fig. \ref{fig:adsr_complete}(a)). The size of the unit cell has a periodicity of $ p=416\,\mu$m. The inner radius is $r=96\,\mu$m, the arc width $ w=20\,\mu$m, the angle $\varphi_{1}=42^\circ$, and the opening angle $\varphi_{2}=22^\circ$. The design is symmetrical to an imaginary line inclined by $45^\circ$ between the x- and y-axis (red line in Fig. \ref{fig:adsr_complete}(a)). The design assumes a certain polarization direction of the incident wave denoted as $ \overrightarrow{E}_{Inc}$ (Fig. \ref{fig:adsr_complete}(a)). The wave propagates in a direction perpendicular to the sensor surface.

Similar structures of gold aDSRRs on glass substrates were presented previously \cite{DebusBolivar2007}. Here, we designed the complementary structure, i.e., aDSRRs as slits in a chro\-mium\-/gold\-/chro\-mium layer on a quartz glass substrate, by applying Babinet's principle \cite{Falcone2004, DebusBolivar2008}. This complementary design has several advantages: (i) aDSRRs are measured in transmission mode, which is easier to realize and handle than reflection mode; (ii) this design allows for additional optimization through an undercut etched into the substrate, a key design feature that results in a higher sensitivity by enabling the selective functionalization of open gold surfaces in those areas where electric field is maximal.

To visualize the layout of the etched undercut, a schematic cross section of the freestanding metallic structure is shown in Fig. \ref{fig:adsr_complete}(b). The E-field is highly confined at the edge of the arcs of the aDSRRs and asymmetric towards the substrate, thereby concentrated underneath the freestanding metal. Figure \ref{fig:adsr_complete}(d) shows the distribution of the electric field of the long arc. The refractive index of the quartz substrate is higher than for surrounding air, making the coupling of the E-field to the glass substrate more efficient. This results in the asymmetry of the E-field towards the substrate. The field enhancement in the lower left edge of Fig. \ref{fig:adsr_complete}(d) is a simulation artefact due to the sharp geometrical shape of the simulation model, which can be neglected, since the etching process during fabrication results in a round shape (cf. Fig. \ref{fig:adsr_complete}(c)). The localized functionalization with capture DNA is depicted in green in Fig. \ref{fig:adsr_complete}(b) as a thin film on the Au surface. Utilizing a freestanding structure to increase sensitivity has a tradeoff with mechanical stability. This fact has to be considered during each handling, cleaning, and preparation step.

The aDSRRs are arranged in groups of 5x5 and 7x7 elements in separated areas (query fields) (Fig. \ref{fig:adsr_complete}(e)). The compact design allows for the configuration of up to 30 query fields, allowing for reference and multiple experiments on one biosensor within a single measurement cycle.

\subsection{Fabrication}
The metallization of the aDSRR structures are fabricated on top of a $500\,\mu$m quartz substrate (Fused Silica UV grade). The THz refractive index of fused silica is $ n\approx 1.96 $ and remains constant in the THz range \cite{Naftaly2007}. The deposited metallization consists of a $h_{Cr1}=10\,$nm chromium layer which is the adhesive agent for a $h_{Au}=200\,$nm gold layer. As surface passivation, an additional $h_{Cr2}=10\,$nm chromium layer is placed on top (cf. Fig. \ref{fig:adsr_complete}(b)). The metal is structured using standard photolithography and wet etching processes. Subsequently, an additional wet etching process with hydrofluoric acid (HF) is applied to make a $d_{Etch}=3\,\mu$m deep isotropic undercut etch ($d_{Etch}=h_{Etch}$) in the quartz substrate. This etching process creates a freestanding metal structure (Fig. \ref{fig:adsr_complete}(c)). The isotropic HF etching profile results in an undercut of $3.08\,\mu$m by $3.33\,\mu$m, which agrees very well with the design objective. The chromium layer underneath the freestanding metal is wet etched to uncover the gold surface at this undercut area only, in order to allow selective biomolecular functionalization of this area. Finally, photoresist is removed and the wafer is cleaned.

\subsection{Modeling results}
To evaluate and optimize the properties of the aDSRR, we used the Maxwell solver ANSYS HFSS for electromagnetic modeling. The aDSRR unit cell was simulated as a repetitive element with periodicity \textit{p} and periodic boundary conditions. Gold was modeled with a dielectric constant $\varepsilon_{r}=-1.12\times 10^{5}$ and conductivity $\sigma=4.01 \times 10^{7}\,$S/m. These parameters were calculated from the complex dielectric function of the Drude model \cite{Johnson1972}. Chromium layers were not modeled, as due to the low thickness of the layer, no influence on the simulation results can be observed. The material properties of quartz glass are $\varepsilon_{r}=3.81$ with a dielectric loss tangent $ \delta=0.0001 $ \cite{Naftaly2007,Grischkowsky1990}. The modeling of the dielectric loading applied through the DNA film is particularly challenging due to the vast difference in DNA molecule size compared to the THz wavelength, and the shortage of reliable data on the dielectric properties of DNA in the THz range.

\begin{figure}[ht]
\centering
\includegraphics[width=0.8\textwidth]{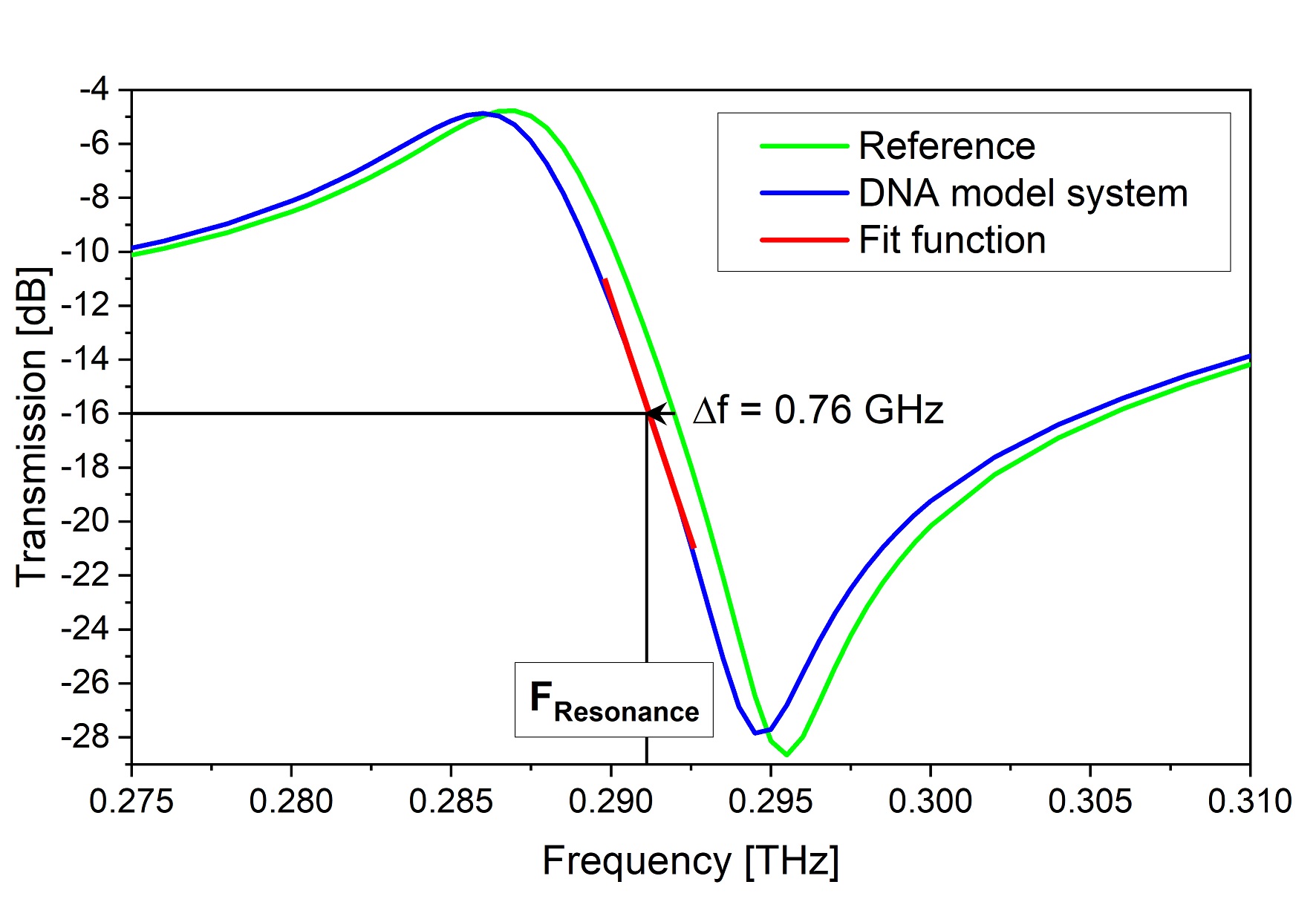}
\caption{Simulated transmission spectra for the designed aDSRR structure of reference and simulated DNA loading. The peak-to-peak transmission intensity difference of both resonance features is $\sim 23\,$dB with an $8.5\,$GHz width. Comparing the center frequencies of both simulations, a shift of $\Delta f=0.76\,$GHz towards lower frequencies is observed as a result of DNA loading.}
\label{fig:simulation}
\end{figure}

Few attempts have been made to determine the physical properties of DNA in the THz frequency range. In \cite{Markelz2000} the refractive index of large DNA with a mass of $\sim\,12\,\times\,10^6\,$g/mol was measured as $ n\leq1.5 $. Brucherseifer et al. \cite{BrucherseiferNagelBolivarEtAl2000} worked out the difference between denatured and hybridized DNA samples for large DNA molecules ($ 5.4\,$kb) in thick layers ($ 30\,\mu$m). It was obtained $ n\approx1.05 $ for denatured and $ n\approx1.175 $ for hybridized DNA at a frequency of $1.25\,$THz. However, the DNA samples were arranged differently in our study: as short DNA strands ($24$ - $25\,$bp) functionalized on a gold surface. The DNA molecules self-assembled on a film but not all binding sites were occupied. Therefore, these samples act in a different manner than dry disordered samples of $30\,\mu$m bulk DNA. Previous experiments with denatured and hybridized DNA on a stripline dipole resonator have shown a frequency shift that is 2600 times larger than expected by simulation results \cite{NagelBolivarBrucherseiferEtAl2002a, Baras2003}. These experiments clearly demonstrate the different behaviour of bulk DNA compared to self-assembled monolayers of DNA on a sensor surface. As a consequence, a comparative simulation and experimental analysis loading the sensors with DNA and with a  polymer film was performed \cite{Nagel2003a}. The resulting frequency shift caused by the dielectric loading with a polymer film is comparable to the shift caused by the self-assembled DNA film. This demonstrates that it is viable to simplify simulations by exchanging the extremely thin DNA film by a $100\,$nm PMMA film \cite{Debus2013}. Our model for the DNA film is approximated as a $100\,$nm thick PMMA layer with the material parameters $\varepsilon_{rDNA}=2.6$, bulk conductivity $\sigma=100\,$kS/m, and dielectric loss tangent $ \delta=0.001 $ \cite{Jin2006}, therefore. This model was verified by comparing simulation results with measurements to ensure that it correctly describes the dielectric loading of DNA on the aDSRR in simulations. In order to determine the shift of the position of the resonance frequency as a result of the dielectric loading with the DNA model system, the DNA (blue curve in Fig. \ref{fig:simulation}) in the simulation model was exchanged with air as a reference (green curve in Fig. \ref{fig:simulation}, $\varepsilon_{rAir}=1$). The shift of the resonance frequency was then calculated by the difference of reference and DNA model system.

The superposition of the resonances of each split ring arc results in a Fano-type double resonance feature (DRF) in the simulated transmission spectrum (Fig. \ref{fig:simulation}). The properties of the DRF are given by the maximum and minimum transmissions, the peak-to-peak transmission intensity difference ($\sim 23\,$dB), width $8.5\,$GHz, and flank steepness $-2.71\,$dB/GHz. The position of the resonance feature was determined through a linear fit function of the resonance flank between $289.8$ and $292.6\,$GHz, which contains the inflection point. We determined the position of the resonance frequency exclusively from the linear part of the resonance feature at $F_{Resonance}=291\,$GHz with the inflection point at $-16\,$dB. This fitting method is essential to determine the position of the resonance frequency at the intersection level at $-16\,$dB, which is in the center between maximum and minimum of the DRF. The linear fitting ensures the robust detection of the resonance frequency and thus a reproducible and repeatable determination of the frequency shift. As a results of the dielectric loading with the DNA model system, its resonance frequency was shifted by about $0.76\,$GHz towards lower frequencies compared to the reference. Please note that the absolute position of the DRF in the measurement results was lower (see Section 3) as a result of the shape variation of the fabrication process in comparison to the simulation model. However, as we are interested in relative frequency shifts, the absolute DRF position is irrelevant. 

\subsection{Measurement setup: biosensor reader}
Traditional THz measurement systems for material analysis and characterization typically use time domain spectroscopy (THz-TDS) consisting of photoconductive antennas for THz generation and electro-optical sampling \cite{Grischkowsky1990,Auston1984,Cheung1985}. These systems generate a frequency spectrum that has a bandwidth of several THz, but limited resolution. Therefore, THz-TDS systems are not suitable to detect small resonance shifts.

We used a robust and frequency-stable all-electronic THz spectroscopy system in transmission mode for our biosensor measurements, therefore. This system provides a signal-to-noise ratio over $40\,$dB and can detect small frequency shifts below $6\,$MHz. The biosensor was positioned by motorized stages with $5\,\mu$m position repeatability in all three directions, allowing a very high measurement reproducibility for consecutive measurements after different biochemical procedures. The all-electronic measurement scheme is based on quasi-heterodyne detection: a voltage-controlled oscillator (VCO), modulated by a continuous saw-tooth signal, generating a local oscillator (LO) frequency of $12-18\,$GHz. The LO signal is multiplied by a factor of 18 in a frequency multiplier chain, which results in an operating frequency range of $220-320\,$GHz. Consequently, measuring one query field takes $110\,$ms, including signal processing. A more detailed description of the setup is provided in \cite{DebusSpickermannNagelEtAl2011}.

\subsection{Sample preparation and methods}
Two kinds of measurements were performed: (1) using \textit{ex-situ} hybridized synthetic DNA to test functionalization dependencies, and (2) using total RNA derived from pathological samples of human origin (including melanoma inhibitory activity (MIA) mRNA) in an \textit{on-chip} hybridization to mimic typical biosensor applications.

Initially, the chips were cleaned of residues, especially organics, with piranha solution (H$_{2}$SO$_{4}$ : H$_{2}$O$_{2}$, 3:1) for $ 2\,$h at $ 75\,^\circ$C. After cleaning, the remaining piranha solution was washed away by intensive rinsing with deionized (DI) water. The chip was dried in a nitrogen flow and placed in an oven at $50\,^\circ$C for $45\,$min.

The functionalization of the capture oligonucleotides is based on thiol-gold binding \cite{LoveEstroffKriebelEtAl2005}. All capture probes were modified at their $5'$-ends with a thiol group, allowing for localized functionalization to the undercut etched gold surface due to the specific binding of thiol to gold. The functional thiol group was protected by mercaptohexanol (MCH), forming a disulfide bond. Prior to use, the disulfide was deprotected with tris(2-carboxyethyl)phosphin (TCEP). After each functionalization step, $1\,$mM MCH solution removed unspecific bindings of adsorbed DNA. The MCH forms a self-assembling monolayer (SAM) and improves the accessibility of the captured oligonucleotides for the target DNA. Therefore, the hybridization efficiency is maximized and the fully covered surface suppresses nonspecific adsorption to the gold surface \cite{HerneTarlov1997}.

Each functionalization and hybridization steps involved manually pipetting $3\,\mu$l sample volumes, forming a droplet, which was left for $45\,$min on the biosensor at room temperature to ensure functionalization and hybridization were successful. After, the biosensor was rinsed with DI water to remove residues. In the last step the biosensor was dried with nitrogen and tempered at $50\,^\circ $C for $45\,$min, followed by biosensor measurement. All measurements were performed in dry state to prevent distortion caused by water molecules. The drying process of the double-stranded (ds) DNA does not affect the bound cDNA, having no influence on results.

For the unequivocal identification of frequency shifts in a query field, each biochemical process step was followed by a complete measurement of the biosensor. The frequency shift of each query field was determined by comparing the positions of the resonance feature of two measurements. The frequency shift on unloaded query fields were used as reference. Reference measurements are vital to control and monitor shifts, which may result from unwanted chemical or mechanical effects.

\subsubsection{Preparation of ex-situ hybridized synthetic DNA}
\textit{Ex-situ}, in this context, is the hybridization process of capture and target DNA in buffer solution. The hybridization process was performed in a tube filled with buffer solution in a thermoshaker for $8\,$h at $50\,^\circ$C, below the melting temperature for this dsDNA ($67.8\,^\circ$C). Once hybridization was complete, separated sulfides were removed in Bio-Spin\textregistered P-6 Gel Columns centrifuged at $1000\,$g for $4\,$min with hybridization buffer ($ 1\,$M NaCl, $ 10\,$mM Tris-HCl, $ 1\,$mM EDTA, and NaOH to adjust to pH $7.5$). Denaturation of dsDNA was performed with $8\,$M solution of CO(NH$_{2}$)$_{2}$ (UREA) for $ 45\,$min.

The $25\,$bp DNA sequence derived from beta-actin mRNA is: $5'$-THI-TGG-CAC-CAC-ACC-TTC-TAC-AAT-GAG-C-$3'$ for the capture and $5'$-G-CTC-ATT-GTA-GAA-GGT-GTG-GTG-CCA-FLU-$3'$ for target probes, hereinafter referred to as SYN1. The $24\,$bp DNA sequence derived from MIA mRNA is: $5'$-THI-GGT-CCT-ATG-CCC-AAG-CTG-GCT-GAC-$3'$ for the capture and $5'$-GTC-AGC-CAG-CTT-GGG-CAT-AGG-ACC-FLU-$3'$ for target probes, hereinafter referred to as SYN2. The fluorescence marker at the $3'$-end of the target probe enables further analysis without affecting THz measurements. The oligonucleotides were ordered from Eurofins Genomics GmbH.

\subsubsection{Preparation of on-chip hybridized copy DNA}
\textit{On-chip} hybridization is the process of cDNA strands hybridizing to the biosensor surface. Functionalization was performed with the thiol modified synthetic capture probe SYN2, described above. This sequence is able to capture one specific cDNA, MIA, of the total cDNA sequences prepared from total RNA isolated from the human melanoma cell line (with the share of MIA cDNA being as low as $0.000048$\%). The specific detection of one of these cDNA molecules covers the typical case of application of medical diagnosis. To measure endogenous DNA molecules, aliquots of $100\,\mu$l containing cDNA from the melanoma cell line Mel Im were produced. Isolation of the total RNA was carried out after pelleting the melanoma cells cultured \textit{in vitro} ($\sim 3$ million cells per pellet). To perform the isolation, the manufacturer's protocol of the E.Z.N.A.\textsuperscript{\textregistered} Total RNA Kit I (Omega Bio-tek) was followed. Subsequently, $500\,$ng RNA was reverse transcribed in a volume of $20\,\mu$l following the manufacturer's protocol of reverse transcriptase SuperScript\texttrademark\,II (Invitrogen AG) at $37\,^\circ$C for $45\,$min using dN$_{6}$ primer. The reverse transcriptase reaction buffer system contains $1\,\mu$l reverse transcriptase II, $4\,\mu$l First-Strand Buffer ($ 250\,$mM Tris-HCl (pH 8.3), $ 375\,$mM KCl, $ 15\,$mM MgCl$_{2}$), $2\,\mu$l $ 0.1\,$M DTT, $1\,\mu$l $10\,\mu$M dNTPs (Genaxxon bioscience GmbH), $1\,\mu$l random primer dN$_{6}$ (Roche Deutschland Holding GmbH) and H$_{2}$O bidest. adding up to $20\,\mu$l. Subsequently, the remaining RNA was digested by adding $1\,\mu$l RNase and incubating for $30\,$min at $37\,^\circ$C. No additional PCR amplification was performed in this study, therefore the generated amount of cDNA reflects 1:1 the endogenous amount of RNA.

The functionalization and measurement process begins with the initial functionalization of the query fields with $20\,\mu$M ssDNA of the synthetic capture SYN2 sequence. The measurement fields were subsequently treated with $1\,$mM MCH for $45\,$min to remove unspecific bound ssDNA. Next, $8\,$M UREA was pipetted on query fields which were initially functionalized with dsDNA (data not shown). Finally, the \textit{on-chip} hybridization occurred by adding the complementary ssDNA sequence of $3\,\mu$l cDNA generated as described above. Each process step was followed by cleaning: DI water rinse, nitrogen drying, and tempering at $50\,^\circ $C for $45\,$min. Measurement of the biosensor followed each step.

\section{Results and discussion}
\subsection{Functionalization dependency test with ex-situ hybridized dsDNA}
To verify functionalization and hybridization processes, the biosensor was tested with the \textit{ex-situ} hybridized dsDNA sequences SYN1 and SYN2. \textit{Ex-situ} dsDNA is particularly convenient because hybridization efficiency in solution is high compared to \textit{on-chip} hybridization. Additionally, it is an established preparation step prior to \textit{on-chip} hybridization.

\begin{figure}[ht]
\centering
\includegraphics[width=\textwidth]{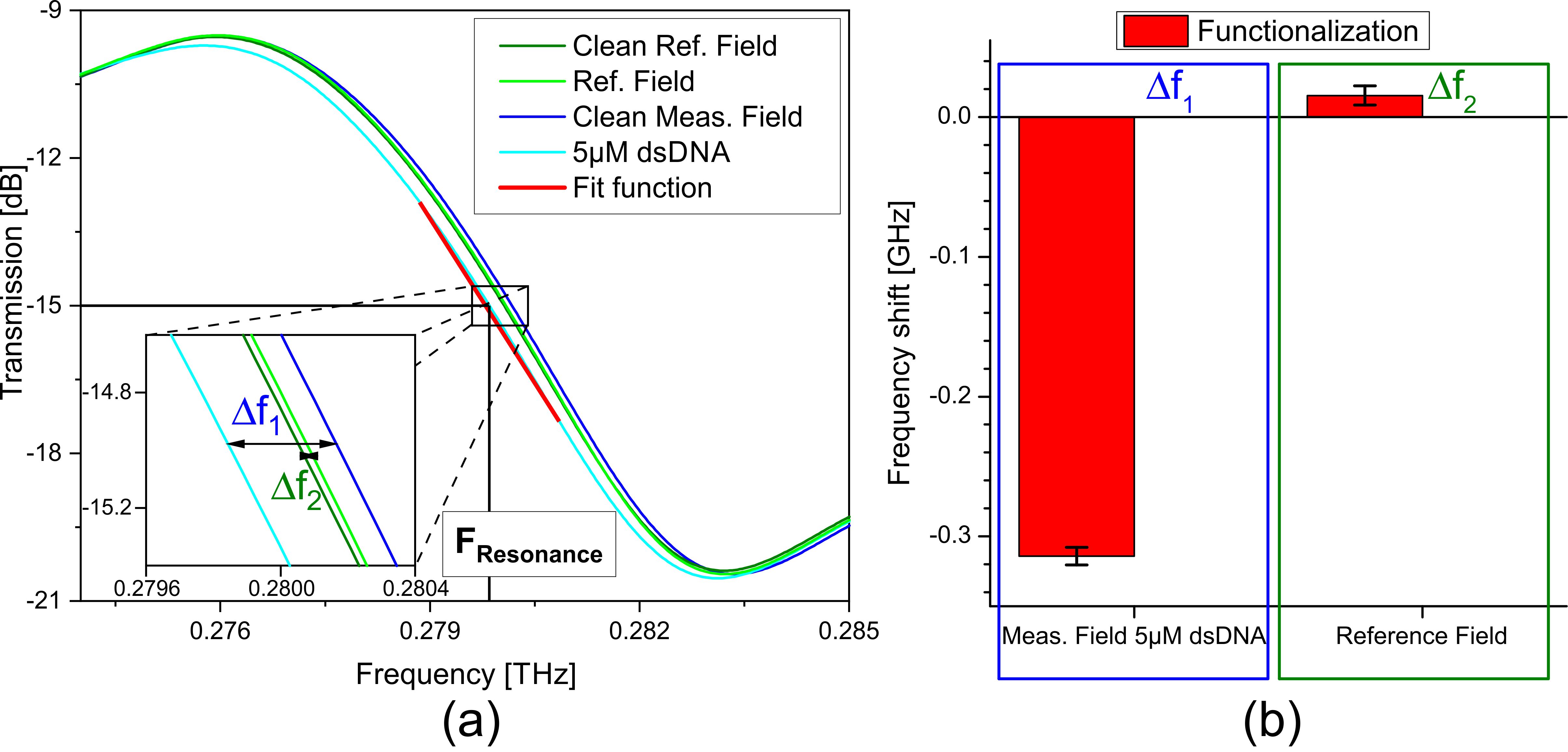}
\caption{(a) Representative transmission spectra of four measurements at two query fields. The resonance frequency $F_{R}$ of each measurement was determined by a fitting and analyzed at $-15\,$dB. A frequency shift of $ \Delta f_{1}=-314\,$MHz is observed for the query field loaded with $5\,\mu$M dsDNA in comparison the frequency shift $\Delta f_{2}$ of the reference field is negligibly small (b).}
\label{fig:data_analysis}
\end{figure}

Prior to the experiment, the biosensor was cleaned with piranha solution. The first measurement is the reference for all measurements, referred to as "Clean Ref. Field" and "Clean Meas. Field", see the transmission spectra in Figure \ref{fig:data_analysis}(a). In the following measurement, one field (green curves) was untreated while the second query field was functionalized with the \textit{ex-situ} hybridized $5\,\mu$M dsDNA sequence SYN1 (blue curves). The DRF on the second query field is shifted by $\Delta f_{1}=314\,$MHz towards lower frequencies in relation to the clean measurement field (cf. in Figure \ref{fig:data_analysis}(a) "Clean Meas. Field" and "$5\,\mu$M dsDNA"). Comparing the measurements "Clean Ref. Field" and "Ref. Field", the frequency shift $\Delta f_{2}$ for the reference field increases by $15\,$MHz. Figure \ref{fig:data_analysis}(b) represents the frequency shifts $\Delta f_{1}$ for the query field functionalized with $5\,\mu$M dsDNA and $\Delta f_{2}$ used as reference. In order to determine statistically relevant data, each biosensor was measured ten times consecutively. These values were used to calculate the mean and experimental standard deviation (maximum $6\,$MHz, see Fig. \ref{fig:data_analysis}(b)).

\begin{figure}[ht]
\centering
\includegraphics[width=0.9\textwidth]{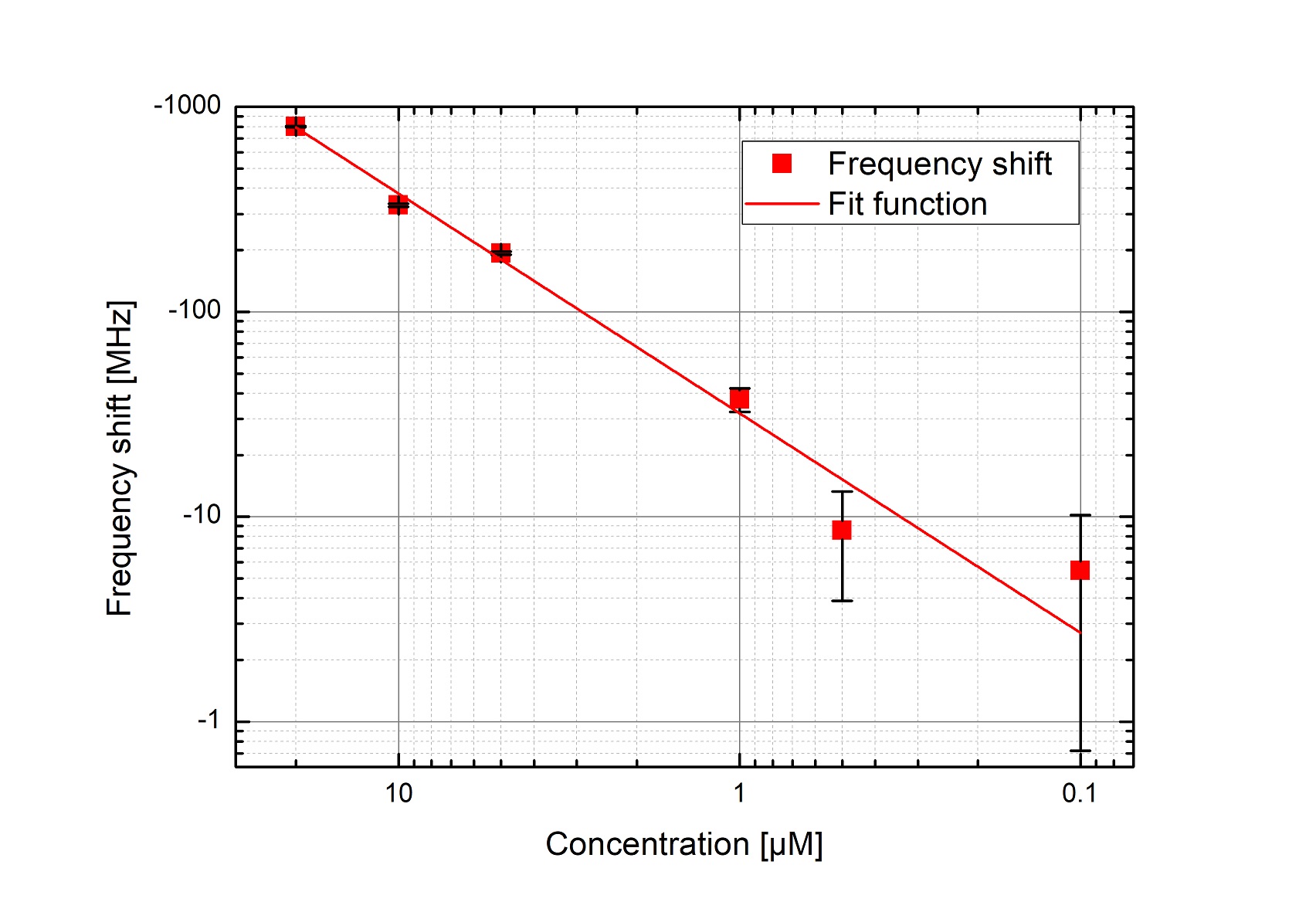}
\caption{Measured frequency shift caused by loading the biosensor with hybridized $24\,$bp dsDNA SYN2. Variation in DNA concentration between $20$ and $0.1\,\mu$M has a dependency related to the frequency shift towards lower resonance frequencies.}
\label{fig:titration}
\end{figure}

The second experiment was a series of measurements using the $24\,$bp dsDNA sequence SYN2 derived from MIA with concentrations varying between $20\,\mu$M and $0.1\,\mu$M. Figure \ref{fig:titration} depicts the measured frequency shift towards lower frequencies dependent on dsDNA concentration in order to assess the selective functionalization efficiency. The experiment was performed on one biosensor in different query fields, which were functionalized with \textit{ex-situ} hybridized dsDNA. The frequency shift reaches from $801\,$MHz for $20\,\mu$M dsDNA concentration to $5.5\,$MHz for $0.1\,\mu$M dsDNA. A distinct correlation between surface occupation and the shift of the position of the resonance frequency is evident. The lowest concentration reaches the lower limit of our detection range and the lowest reliable detectable \textit{ex-situ} hybridized dsDNA concentration is $1\,\mu$M. Please note that this variation of the density thiol group containing dsDNA does measure the functionalization efficiency and not the application relevant \textit{on-chip} hybridization detection concentration requirements, as shown below.

\subsection{Application evaluation with on-chip hybridized cDNA}
\begin{figure}[ht]
\centering
\includegraphics[width=\textwidth]{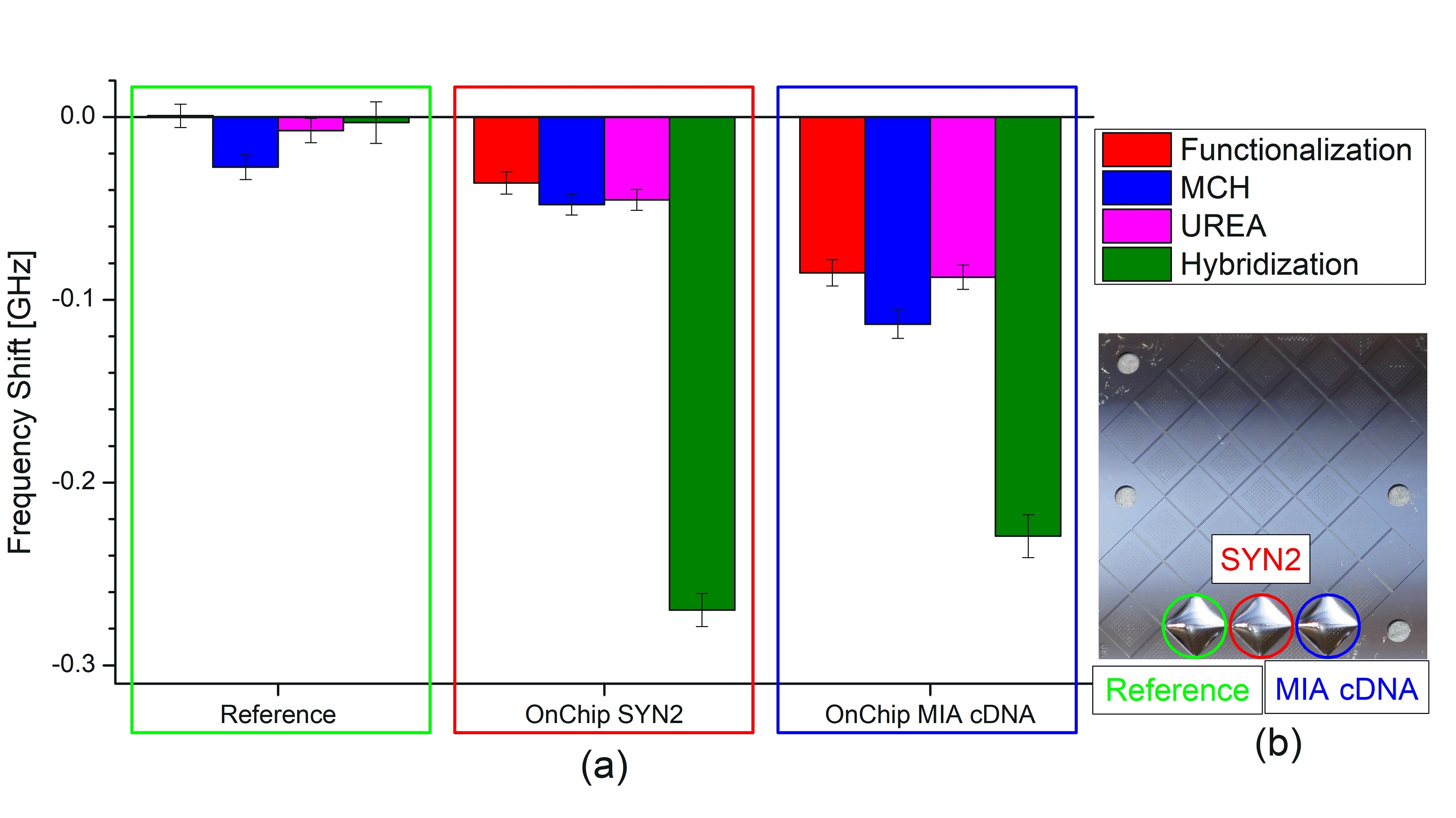}
\caption{(a) Relative frequency shifts towards lower frequencies as a consequence of different DNA sequences which are functionalized on different query fields (indicated by green, red, and blue boxes) on one biosensor. The colored bars indicate the process steps: red is the functionalization with thiol modified capture DNA, blue the treatment with MCH, pink rinsing with UREA, and green the \textit{on-chip} hybridization with the target sequences SYN2 and MIA cDNA. The biosensor loaded with various analytes is illustrated in (b).}
\label{fig:MIA_shift}
\end{figure}

\textit{On-chip} hybridization experiments were performed to demonstrate the application relevant levels for dsDNA detection. Experiments with the synthetic DNA sequence SYN2 and with human melanoma cell line MIA cDNA sample produced from total RNA (without PCR-amplification) are carried out. The amount of MIA cDNA in the sample is as low as $4.64 \times 10^{-18}\,$mol with a molecular weight of $7378\,$g/mol. This amount was calculated by determining the share of MIA mRNA in the isolated total RNA samples using RNA-Seq data sets of the melanoma cell line Mel Im as $0.000048$\% and the amount of total RNA used in the experiment. The measurement results are shown in Figure \ref{fig:MIA_shift}(a). The colored boxes define different query fields, whereby the green box is the untreated reference field, the red is the \textit{on-chip} hybridization with $20\,\mu$M synthetic SYN2 sequence, and the blue box is the \textit{on-chip} hybridization with $1.55 \times 10^{-12}\,$mol/l PCR-free MIA cDNA sample. Fig. \ref{fig:MIA_shift}(b) depicts the biosensor query fields loaded with various samples.

The red bars in Figure \ref{fig:MIA_shift}(a) are the frequency shifts as a result of the functionalization, compared to the initial measurement of the query field. The resonance frequency on the reference field remains at the same position, compared to the fields which were functionalized: a shift of $-36\,$MHz on the SYN2 field and $-85\,$MHz on the MIA cDNA field. The blue bars (MCH treatment) depict an additional slight shift towards lower resonance frequencies when compared to the functionalized case. Since the same shift appears on all measured fields, including the reference field, it must be due to mechanical and chemical handling. UREA was used to denature dsDNA and had no influence on the ssDNA, resulting in negligible changes in resonance frequency. The frequency shift from the hybridization process is shown in green, yielding to $-270\,$MHz for the SYN2 sequence and $-230\,$MHz for MIA cDNA, relative to the initial measurement of the biosensor. Compared to the functionalized resonance frequency (blue bars), the shift is $-234\,$MHz and $-145\,$MHz for SYN2 and MIA cDNA, respectively.

\subsection{Discussion}
The transmission spectra of simulation and experimental results are consistent. The peak-to-peak transmission intensity differences of the measured aDSRR structure is $10\,$dB over a frequency range of $7\,$GHz. Compared to simulations, the peak-to-peak transmission intensity is lower, as a result of the finite array, the losses from the materials and limited conductivity of gold and chromium. However, the width and relative frequency shift of the resonance feature are in agreement with model results.

The first experiments are performed to verify the efficiency of the functionalization procedures. Experimental functionalization dependencies with the \textit{ex-situ} hybridized $25\,$bp dsDNA SYN1 and the $24\,$bp dsDNA SYN2 sequence, show a distinct frequency shift, while the simultaneously measured reference field displays no significant shift. The correlation between frequency shift and dsDNA concentration clearly demonstrates the capability of the biosensor to detect short dsDNA strands at various concentrations, down to a limit of $1\,\mu$M for \textit{ex-situ} hybridized dsDNA. The efficiency of \textit{ex-situ} hybridization in these experiments is at its maximum due to the long interaction time span, controlled temperature, and mixture with buffer solution. The detectable frequency shift is therefore only limited by the number of functionalized capture DNA molecules, which are bound to the gold surface and is directly proportional to dsDNA concentration. Higher absolute shift levels would be expected. The reason for the limitation in the functionalization process is not fully understood since the query fields are loaded with similar probes and processed consecutively. Conceivable reasons are that the single query fields were not fully cleaned of residues from the fabrication. Furthermore, the efficiency of the functionalization process is unknown, since the gold surface on the bottom of the freestanding metal is hard to access for the thiol-modified DNA. We also assume that the texture of the gold surface reduces the functionalization density and the efficiency of the process. Therefore, the functionalization process needs to be optimized for improved reproducibility. However, the experiments verify the functionalization and hybridization capability of the sensor fabrication process and demonstrate the selective functionalization efficiency of our biosensor.

The second experiment represents an application evaluation with pathological samples produced from the human melanoma cell line Mel Im. The \textit{on-chip} hybridization experiment was performed with the synthetic DNA SYN2 and the MIA cDNA as target probes, hybridized on the biosensor surface. The shift caused by the \textit{on-chip} hybridization with MIA cDNA is remarkable, since the MIA cDNA shift reaches $62$\% in comparison to the synthetic SYN2 results. This is particularly notable in view of future THz biosensor applications since the concentration of MIA cDNA is seven orders of magnitude smaller than the one of the synthetic SYN2 sequence. This clearly indicates a close to perfect \textit{on-chip} hybridization efficiency (on the order of $62$\%) and a limited efficiency of the functionalization process. The absolute number of capture DNA molecules, which are available for the \textit{on-chip} hybridization process, is equivalent to the number of dsDNA molecules during the \textit{ex-situ} hybridization experiments. The results indicate that we reached a saturation level in the measurements with \textit{ex-situ} hybridized synthetic SYN2, whereas in the \textit{on-chip} hybridization, the ratio of available capture molecules to applied MIA cDNA allows for an almost equivalent frequency shift. The number of hybridized molecules of MIA cDNA and synthetic SYN2 is therefore almost the same despite the lower concentration of MIA cDNA ($1.55 \times 10^{-12}\,$mol/l).

\begin{figure}[ht]
\centering
\includegraphics[width=\textwidth]{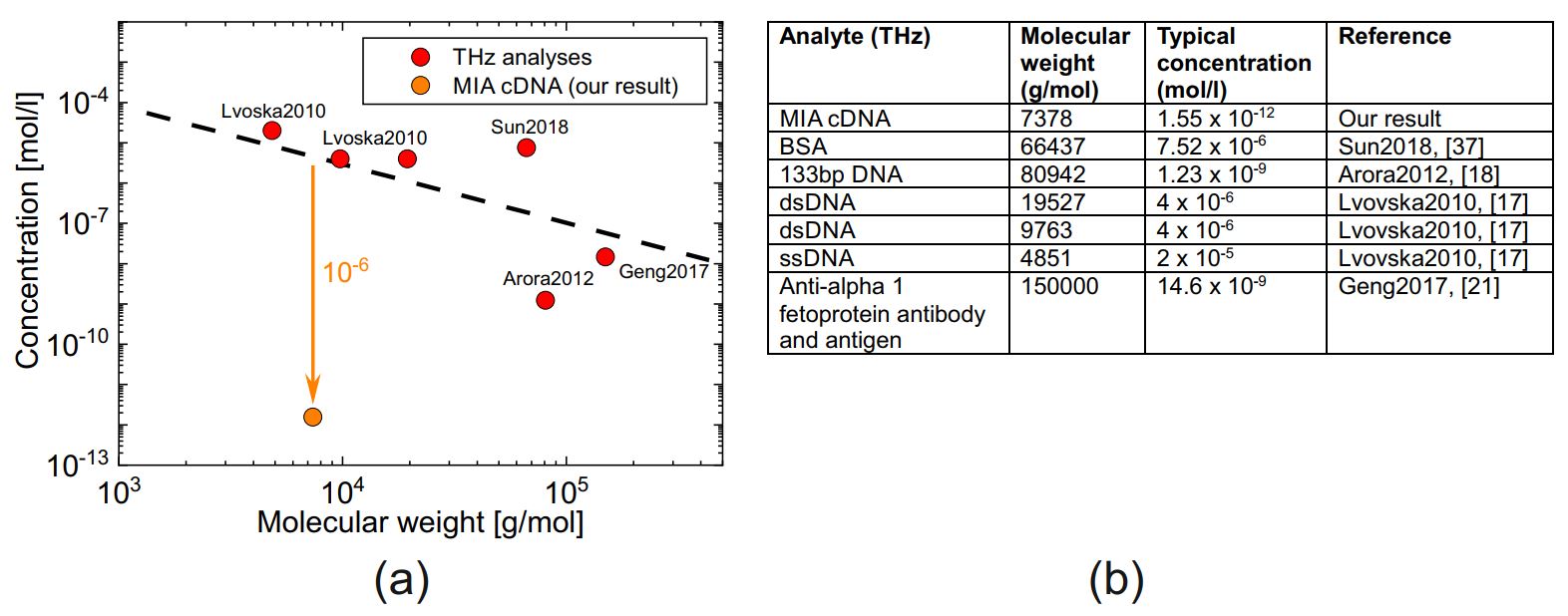}
\caption{(a) In comparison to published THz results, our measurements show increased sensitivity by six orders of magnitude at comparable molecular weights. The dashed line represents the overall trend of decreasing detection limits with increasing molecular weight. (b) Table with best of class THz analyses of biomolecules.}
\label{fig:sensitivity}
\end{figure}

In comparison to state-of-the-art research results of biosensing in the THz range, the detection of MIA cDNA at a PCR-free concentration of $1.55 \times 10^{-12}\,$mol/l is remarkable. The total amount of molecules is as low as $4.64 \times 10^{-18}\,$mol of MIA cDNA, with a droplet size of $3\,\mu$l. Fig. \ref{fig:sensitivity}(a) shows a collection of best of class THz measurements of biomolecules (red dots) with different MW indicating the lower detection limit of these analyses \cite{Geng2017,Sun2018,Arora2012,Lvovska2010}. The dashed line in Fig. \ref{fig:sensitivity}(a) indicates the dependency of the minimum detectable concentration on MW with the slope of $m = -1.45$ \cite{Weisenstein2018}. A clear trend is visible: the higher MW, the lower the detectable concentration of biomolecules. 

Our result for PCR-free MIA cDNA reaches a sensitivity roughly six orders of magnitude better than existing approaches for biomolecules with comparable MW. On top of the high sensitivity, our approach is highly specific to the target DNA by utilizing the highly selective hybridization process of cDNA strands. Specificity, achieved by functionalization of the sensor surface using capture probe molecules (i.e. ss oligonucleotides), which are chemically bound to the biosensor surface, is a second key feature for bioanalytical applications addressed with our biosensor.

\section{Conclusion}
In this work, we present the development, fabrication, and practical use of a THz biosensor for the tumor-related analysis of DNA samples. Based on electromagnetic field simulations, the design was optimized to maximize sensitivity through selective functionalization, undercut etched structures, and resonance of the split ring resonators. The resonance frequency was determined with an accuracy of $6\,$MHz in an all-electronic spectroscopic system at a center frequency of $280\,$GHz. The use of the biosensor for the detection of synthetic and human DNA samples was experimentally verified by \textit{ex-situ} and \textit{on-chip} hybridization in different concentrations which were reproduced in model simulations. We detected a frequency shift of $230\,$MHz for MIA cDNA and $270\,$MHz for a synthetically produced DNA sequence. Highly specific hybridization of MIA cDNA enables the detection out of a vast number of various DNA molecules, which mimics the typical case of real-world applications for medical diagnosis. Most importantly, the experimentally determined lower detection limit is $1.55 \times 10^{-12}\,$mol/l for MIA cDNA without PCR amplification, equal to $4.64 \times 10^{-18}\,$mol of molecules. These results are by six orders of magnitude better than published THz results. With the demonstrated exceptionally high PCR-free sensitivity combined with a very high specificity, achieved by biochemical sensor functionalization, we addressed two key features for biosensor application. Our biosensor outperforms existing approaches in two essential aspects: PCR-free sensitivity and specificity, paving the way for a broader implementation and use of THz biosensing techniques.

\section*{Funding}
Deutsche Forschungsgemeinschaft (BO 1573/27-1, BO 1573/27-2, HA 3022/8-1, HA 3022/8-2,
SCH 1970/1-1, WI 5209/1-2).

\section*{Acknowledgment}
This work is part of the national priority program SPP 1857 ESSENCE.

\section*{Disclosures}
The authors declare that there are no conflicts of interest related to this article.


\bibliography{PCRpaper}

\begin{thebibliography}{10}
\newcommand{\enquote}[1]{``#1''}

\bibitem{Vo-Dinh2001}
T.~Vo-Dinh, B.~M. Cullum, and D.~L. Stokes, \enquote{Nanosensors and biochips:
  frontiers in biomolecular diagnostics,} {\protect\JournalTitle{Sensors and
  Actuators B: Chemical}} \textbf{74}, 2--11 (2001).

\bibitem{Liu2004}
R.~H. Liu, J.~Yang, R.~Lenigk, J.~Bonanno, and P.~Grodzinski,
  \enquote{Self-contained, fully integrated biochip for sample preparation,
  polymerase chain reaction amplification, and {DNA} microarray detection,}
  {\protect\JournalTitle{Analytical chemistry}} \textbf{76}, 1824--1831 (2004).

\bibitem{Wang2000}
J.~Wang, \enquote{Survey and summary: from {DNA} biosensors to gene chips,}
  {\protect\JournalTitle{Nucleic acids research}} \textbf{28}, 3011--3016
  (2000).

\bibitem{Ozaki1992}
H.~Ozaki and L.~W. McLaughlin, \enquote{The estimation of distances between
  specific backbone-labeled sites in {DNA} using fluorescence resonance energy
  transfer,} {\protect\JournalTitle{Nucleic acids research}} \textbf{20},
  5205--5214 (1992).

\bibitem{Zhu1994}
Z.~Zhu, J.~Chao, H.~Yu, and A.~S. Waggoner, \enquote{Directly labeled {DNA}
  probes using fluorescent nucleotides with different length linkers,}
  {\protect\JournalTitle{Nucleic acids research}} \textbf{22}, 3418--3422
  (1994).

\bibitem{Zhu1997}
Z.~Zhu and A.~Waggoner, \enquote{{Molecular mechanism controlling the
  incorporation of fluorescent nucleotides into {DNA} by {PCR}},}
  {\protect\JournalTitle{{CYTOMETRY}}} \textbf{{28}}, {206--211} ({1997}).

\bibitem{Larramendy1998}
M.~L. Larramendy, W.~El-Rifai, and S.~Knuutila, \enquote{Comparison of
  fluorescein isothiocyanate-and texas red-conjugated nucleotides for direct
  labeling in comparative genomic hybridization,}
  {\protect\JournalTitle{Cytometry Part A}} \textbf{31}, 174--179 (1998).

\bibitem{Froehlich1968}
H.~Froehlich, \enquote{Long-range coherence and energy storage in biological
  systems,} {\protect\JournalTitle{International Journal of Quantum Chemistry}}
  \textbf{2}, 641--649 (1968).

\bibitem{Zhuang1990}
W.~Zhuang, Y.~Feng, and E.~W. Prohofsky, \enquote{Self-consistent calculation
  of localized {DNA} vibrational properties at a double-helix-single-strand
  junction with anharmonic potential,} {\protect\JournalTitle{Phys. Rev. A}}
  \textbf{41}, 7033--7042 (1990).

\bibitem{BrucherseiferNagelBolivarEtAl2000}
M.~Brucherseifer, M.~Nagel, P.~H. Bol{\'\i}var, H.~Kurz, A.~Bosserhoff, and
  R.~B{\"u}ttner, \enquote{Label-free probing of the binding state of {DNA} by
  time-domain terahertz sensing,} {\protect\JournalTitle{Applied Physics
  Letters}} \textbf{77}, 4049--4051 (2000).

\bibitem{Markelz2000}
A.~Markelz, A.~Roitberg, and E.~J. Heilweil, \enquote{Pulsed terahertz
  spectroscopy of {DNA}, bovine serum albumin and collagen between 0.1 and 2.0
  {THz},} {\protect\JournalTitle{Chemical Physics Letters}} \textbf{320},
  42--48 (2000).

\bibitem{Mickan2002}
S.~P. Mickan, A.~Menikh, H.~Liu, C.~A. Mannella, R.~MacColl, D.~Abbott,
  J.~Munch, and X.-C. Zhang, \enquote{Label-free bioaffinity detection using
  terahertz technology,} {\protect\JournalTitle{Physics in Medicine \&
  Biology}} \textbf{47}, 3789 (2002).

\bibitem{Fischer2002}
B.~Fischer, M.~Walther, and P.~U. Jepsen, \enquote{Far-infrared vibrational
  modes of {DNA} components studied by terahertz time-domain spectroscopy,}
  {\protect\JournalTitle{Physics in Medicine \& Biology}} \textbf{47}, 3807
  (2002).

\bibitem{DebusBolivar2007}
C.~Debus and P.~H. Bol{\'\i}var, \enquote{Frequency selective surfaces for high
  sensitivity terahertz sensing,} {\protect\JournalTitle{Applied Physics
  Letters}} \textbf{91}, 184102 (2007).

\bibitem{Fedotov2007a}
V.~A. Fedotov, M.~Rose, S.~L. Prosvirnin, N.~Papasimakis, and N.~I. Zheludev,
  \enquote{Sharp trapped-mode resonances in planar metamaterials with a broken
  structural symmetry,} {\protect\JournalTitle{Phys. Rev. Lett.}} \textbf{99},
  147401 (2007).

\bibitem{DebusAwadNagelEtAl2009}
C.~Debus, M.~Awad, M.~Nagel, and P.~H. Bol{\'\i}var, \enquote{Terahertz biochip
  technology: Toward high-sensitivity label-free {DNA} sensors,}
  {\protect\JournalTitle{Am. Biotechnol. Lab}} \textbf{27}, 8--11 (2009).

\bibitem{Lvovska2010}
M.~I. Lvovska, N.~C. Seeman, R.~Sha, T.~R. Globus, T.~B. Khromova, and T.~S.
  Dorofeeva, \enquote{{THz} characterization of {DNA} four-way junction and its
  components,} {\protect\JournalTitle{IEEE Transactions on Nanotechnology}}
  \textbf{9}, 610--617 (2010).

\bibitem{Arora2012}
A.~Arora, T.~Q. Luong, M.~Kr{\"u}ger, Y.~J. Kim, C.-H. Nam, A.~Manz, and
  M.~Havenith, \enquote{Terahertz-time domain spectroscopy for the detection of
  {PCR} amplified {DNA} in aqueous solution,} {\protect\JournalTitle{Analyst}}
  \textbf{137}, 575--579 (2012).

\bibitem{Laurette2012}
S.~Laurette, A.~Treizebre, A.~Elagli, B.~Hatirnaz, R.~Froidevaux, F.~Affouard,
  L.~Duponchel, and B.~Bocquet, \enquote{Highly sensitive terahertz
  spectroscopy in microsystem,} {\protect\JournalTitle{Rsc Advances}}
  \textbf{2}, 10064--10071 (2012).

\bibitem{ELISA179887}
{Company Abcam plc}, \enquote{Human albumin elisa kit (ab179887),}  (2019).

\bibitem{Geng2017}
Z.~Geng, X.~Zhang, Z.~Fan, X.~Lv, and H.~Chen, \enquote{A route to terahertz
  metamaterial biosensor integrated with microfluidics for liver cancer
  biomarker testing in early stage,} {\protect\JournalTitle{Scientific
  reports}} \textbf{7}, 16378 (2017).

\bibitem{Falcone2004}
F.~Falcone, T.~Lopetegi, M.~Laso, J.~Baena, J.~Bonache, M.~Beruete,
  R.~Marqu{\'e}s, F.~Mart{\'\i}n, and M.~Sorolla, \enquote{Babinet principle
  applied to the design of metasurfaces and metamaterials,}
  {\protect\JournalTitle{Physical review letters}} \textbf{93}, 197401 (2004).

\bibitem{DebusBolivar2008}
C.~Debus and P.~H. Bol{\'\i}var, \enquote{Terahertz biosensors based on double
  split ring arrays,} in \emph{Photonics Europe,}  (International Society for
  Optics and Photonics, 2008), pp. 69870U--69870U.

\bibitem{Naftaly2007}
M.~Naftaly and R.~E. Miles, \enquote{Terahertz time-domain spectroscopy for
  material characterization,} {\protect\JournalTitle{Proceedings of the IEEE}}
  \textbf{95}, 1658--1665 (2007).

\bibitem{Johnson1972}
P.~B. Johnson and R.~W. Christy, \enquote{Optical constants of the noble
  metals,} {\protect\JournalTitle{Phys. Rev. B}} \textbf{6}, 4370--4379 (1972).

\bibitem{Grischkowsky1990}
D.~Grischkowsky, S.~Keiding, M.~Van~Exter, and C.~Fattinger,
  \enquote{Far-infrared time-domain spectroscopy with terahertz beams of
  dielectrics and semiconductors,} {\protect\JournalTitle{JOSA B}} \textbf{7},
  2006--2015 (1990).

\bibitem{NagelBolivarBrucherseiferEtAl2002a}
M.~Nagel, P.~H. Bol{\'\i}var, M.~Brucherseifer, H.~Kurz, A.~Bosserhoff, and
  R.~B{\"u}ttner, \enquote{Integrated {THz} technology for label-free genetic
  diagnostics,} {\protect\JournalTitle{Applied Physics Letters}} \textbf{80},
  154--156 (2002).

\bibitem{Baras2003}
T.~Baras, T.~Kleine-Ostmann, and M.~Koch, \enquote{On-chip {THz} detection of
  biomaterials: A numerical study,} {\protect\JournalTitle{Journal of
  Biological Physics}} \textbf{29}, 187--194 (2003).

\bibitem{Nagel2003a}
M.~Nagel, F.~Richter, P.~Haring~Bol{\'\i}var, and H.~Kurz, \enquote{A
  functionalized {THz} sensor for marker-free {DNA} analysis,}
  {\protect\JournalTitle{Physics in Medicine \& Biology}} \textbf{48}, 3625
  (2003).

\bibitem{Debus2013}
C.~Debus, \enquote{A high-sensitivity {THz}-sensing technology for {DNA}
  detection with split-ring resonator based biochips,} Ph.D. thesis, Siegen,
  Universit{\"a}t Siegen, Diss., 2013 (2013).

\bibitem{Jin2006}
Y.-S. Jin, G.-J. Kim, and S.-G. Jeon, \enquote{Terahertz dielectric properties
  of polymers,} {\protect\JournalTitle{Journal of the Korean Physical Society}}
  \textbf{49}, 513--517 (2006).

\bibitem{Auston1984}
D.~H. Auston, K.~Cheung, J.~Valdmanis, and D.~Kleinman, \enquote{Cherenkov
  radiation from femtosecond optical pulses in electro-optic media,}
  {\protect\JournalTitle{Physical Review Letters}} \textbf{53}, 1555 (1984).

\bibitem{Cheung1985}
K.~Cheung and D.~Auston, \enquote{Excitation of coherent phonon polaritons with
  femtosecond optical pulses,} {\protect\JournalTitle{Physical review letters}}
  \textbf{55}, 2152 (1985).

\bibitem{DebusSpickermannNagelEtAl2011}
C.~Debus, G.~Spickermann, M.~Nagel, and P.~H. Bol{\'\i}var,
  \enquote{All-electronic terahertz spectrometer for biosensing,}
  {\protect\JournalTitle{Microwave and Optical Technology Letters}}
  \textbf{53}, 2899--2902 (2011).

\bibitem{LoveEstroffKriebelEtAl2005}
J.~C. Love, L.~A. Estroff, J.~K. Kriebel, R.~G. Nuzzo, and G.~M. Whitesides,
  \enquote{Self-assembled monolayers of thiolates on metals as a form of
  nanotechnology,} {\protect\JournalTitle{Chemical Reviews}} \textbf{105},
  1103--1170 (2005).

\bibitem{HerneTarlov1997}
T.~M. Herne and M.~J. Tarlov, \enquote{Characterization of {DNA} probes
  immobilized on gold surfaces,} {\protect\JournalTitle{Journal of the American
  Chemical Society}} \textbf{119}, 8916--8920 (1997).

\bibitem{Sun2018}
Y.~Sun, P.~Du, X.~Lu, P.~Xie, Z.~Qian, S.~Fan, and Z.~Zhu,
  \enquote{Quantitative characterization of bovine serum albumin thin-films
  using terahertz spectroscopy and machine learning methods,}
  {\protect\JournalTitle{Biomed. Opt. Express}} \textbf{9}, 2917--2929 (2018).

\bibitem{Weisenstein2018}
C.~{Weisenstein}, D.~{Schaar}, M.~{Schmeck}, A.~K. {Wigger}, A.~K.
  {Bosserhoff}, and P.~H. Bol{\'\i}var, \enquote{Detection of human tumor
  markers with {THz} metamaterials,} in \emph{2018 43rd International
  Conference on Infrared, Millimeter, and Terahertz Waves (IRMMW-THz),}
  (2018), pp. 1--2.

\end{thebibliography}

\end{document}